# Chiral exceptional point and coherent suppression of backscattering in silicon microring with low loss Mie scatterer

Hwaseob Lee[1], Ali Kecebas[2], Feifan Wang[1], Lorry Chang[1], Sahin K. Özdemir[2,3*], Tingyi Gu[1*]

[1] Department of Electrical and Computer Engineering, University of Delaware, Newark, Delaware, 19716, USA

[2] Department of Engineering Science and Mechanics, Pennsylvania State University, University Park, PA 16802 USA

[3] Materials Research Institute, Pennsylvania State University, University Park, PA 16802, USA

* Email: tingyigu@udel.edu, sko9@psu.edu

**Abstract:** Non-Hermitian systems with their spectral degeneracies known as exceptional points (EPs) have been explored for lasing, controlling light transport, and enhancing a sensor's response. A ring resonator can be brought to an EP by controlling the coupling between its frequency degenerate clockwise and counterclockwise traveling modes. This has been typically achieved by introducing two or more nanotips into the resonator's mode volume. While this method provides a route to study EP physics, the basic understanding of how the nanotips' shape and size symmetry impact the system's non-Hermicity is missing, along with additional loss from both in-plane and out-of-plane scattering. The limited resonance stability poses a challenge for leveraging EP effects for switches or modulators, which requires stable cavity resonance and fixed laser-cavity detuning. Here we use lithographically defined asymmetric and symmetric Mie scatterers, which enable subwavelength control of wave transmission and reflections without deflecting to additional radiation channels. We show that those pre-defined Mie scatterers can bring the system to an EP without post tuning, as well as enable chiral light transport within the resonator. Counterintuitively, the Mie scatterer results in enhanced quality factor measured on the transmission port, through coherently suppressing the backscattering from the waveguide surface roughness. The proposed device platform enables pre-defined chiral light propagation and backscattering-free resonances, needed for various applications such as frequency combs, solitons, sensing, and other nonlinear optical processes such as photon blockade, and regenerative oscillators.

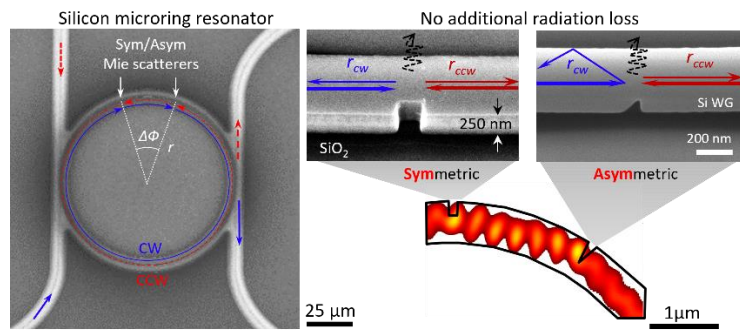

**Key works:** non-Hermicity, exceptional point, Mie scatterers, microring resonator, silicon photonics.

**Introduction**

The past decade has witnessed exciting progress in novel functions and processes enabled by the rich physics and the toolbox provided by non-Hermiticity, especially by the emergence of non-Hermitian spectral degeneracies known as exceptional points (EPs) [1-6]. EPs, where both the eigenvalues and the associated eigenvectors of a system coalesce, are radically different from the degeneracies in Hermitian systems, often referred to as diabolic points (DPs), where eigenvectors are orthogonal even when the eigenvalues coalesce. EPs universally occur in all open physical systems (i.e., non-Hermitian) and dramatically affect their behavior, leading to many counterintuitive phenomena such as loss-induced suppression and revival of lasing [7-8], single-mode lasing [9], directional and mode-selective lasing [10-11], enhanced response to perturbations [11-14], unidirectional invisibility [15], topological control [16-17], and chiral perfect absorption [18], just to name a few.

A physical system can be brought to an EP through judicious engineering of the spatial and spectral distribution of loss and/or gain and of the coupling among constituents of the physical system. Non-Hermiticity and the physics associated with EPs cover all physical systems which exchange energy, information, and mass among themselves or with their environments. In coupled optical waveguides and resonators, tuning the coupling strength between the couples with balanced optical gain and loss (i.e., parity-time symmetric systems) [6] and tuning the loss-imbalance of the couples with fixed coupling strength [7-8] are two commonly used approaches to bring an optical system to an EP. Another approach that has recently gained more attention is controlling the coupling between the modes of a physical system, such as the frequency-degenerate clockwise (CW) and counterclockwise (CCW) modes in a microring resonator (MRR) or a ring laser [19]. This has been realized using nanotip perturbations [11, 15, 19], Taiji resonators (i.e., resonators with an S-shaped inset) [20], or off-the-chip fiber-loop mirrors [21]. In this paper, we use lithographically defined subwavelength dielectric structures, like individual meta-atoms in metamaterials [22-23], etched in the same step as MRRs and grating couplers (see **Fig. 1a, b**). This integrated photonics approach allows us to precisely control the location, size, and geometry of the Mie scatterers, and thus the amplitude, phase and direction of their transmission and reflections. The Mie scatterers embedded in the waveguide are deep subwavelength structures and designed for small reflection (symmetric shape) or deflection at a small angle (asymmetric shape). The energy flow is preserved in the guided modes, and the quality factor ($Q$) degradation is trivial compared to the nanotip perturbation method and Taiji resonators [20]. Here we show that having a Mie scatterer pair formed with one symmetric and one asymmetric Mie scatterer is sufficient to achieve EP or selected chirality in an MRR. The spatial symmetry of the Mie scatterer maps to the different reflection coefficients for CW and CCW modes. With the Mie scatterers in place, we not only achieve EP but also coherently suppress backscattering in ring resonators caused by the fabrication resulted Rayleigh scattering.

**Results**

The non-Hermitian system that we consider in this study is an MRR with two lithographically defined Mie scatterers whose reflection coefficients $r_{1cw(ccw)}$ and $r_{2cw(ccw)}$ mediate the intermodal coupling between the modes [e.g., light in the CW (CCW) mode is reflected into the CCW (CW) mode with $r_{1cw}$ ($r_{1ccw}$) due to the first Mie scatterer]. Three-dimensional numerical simulation provides direct correlation between the reflection coefficients $r_{1cw(ccw)}$ and the Mie scatterer geometry (Fig. S1). For a relatively small depth ($D$ = 100 nm), the phase of reflection coefficients $r_{1cw}$ ($r_{1ccw}$) is proportional to the notch width ($W$) [22] ($D$ and $W$ are marked in **Fig. 1a-b**). Here, this parameter $D$ is carefully selected for introducing sufficient reflection for EP physics studies, but not too large to degrade the transmission $Q$. The distance $L$ between the Mie scatterers leads to a relative phase delay: $\Delta\varphi = 2\pi n_{eff} L/\lambda$ where $\lambda$ denotes the wavelength of the laser wavelength, $n_{eff}$ corresponding to the effective index of the single mode waveguide. A Mie scatterer in the form of a rectangular (**Fig. 1a**) has symmetric scattering in the sense that CW to CCW and CCW to CW reflection coefficients are the same (i.e., $r_{cw} = r_{ccw}$). On the other hand, a Mie scatterer in the form of a right triangle (**Fig. 1b**) defines an asymmetric one where CW to CCW scattering is different than the CCW to CW scattering (i.e., $r_{cw} \neq r_{ccw}$). Also, our numerical studies with practical parameters from the literature show that nanofabrication resulted surface roughness in single mode waveguide leads to asymmetric reflection (Fig. S2).

Our theoretical analysis shows that two embedded Mie scatterers provide sufficient flexibility to tune the system towards or away from its spectral degeneracies. Only one symmetric Mie scatterer is sufficient to counterbalance the backscattering due to fabrication induced Rayleigh scattering. One can write a general Hamiltonian for such a system (Supplementary Section 2):

$$H_2 = \begin{pmatrix} \Omega_0 + \epsilon_{111} + \epsilon_{211} & \epsilon_{112} + \epsilon_{212}e^{-j\Delta\varphi} \\ \epsilon_{121} + \epsilon_{221}e^{j\Delta\varphi} & \Omega_0 + \epsilon_{122} + \epsilon_{222} \end{pmatrix} = \begin{pmatrix} \Omega_0 + \chi_1 & \chi_{12} \\ \chi_{21} & \Omega_0 + \chi_2 \end{pmatrix} \quad (1)$$

whose complex-valued elements $\epsilon_{ikm}$ describes the coupling strength of the $i$-th notch-induced scattering into the same ($k = m$) or the counterpropagating ($k \neq m$) mode. The diagonal complex elements $\epsilon_{i11}$ and $\epsilon_{i22}$ determine the frequency shift and the additional dissipation induced by the Mie scatterers to the CW and CCW modes, respectively. The off-diagonal elements $\epsilon_{i12}$ and $\epsilon_{i21}$ determine the Mie scatterer induced coupling between CW and CCW modes. Thus, without loss of generality, $\epsilon_{1km}$ can be treated as real-valued $\epsilon_{1km} = \epsilon_{1kmr} + i\epsilon_{1kmi}$ where $|\epsilon_{1kmi}| \ll |\epsilon_{1kmr}|$) [22]. The eigenvalues of $H_2$ are $\omega_\pm = \Omega_0 + (\chi_{21} + \chi_{21})/2 \pm \xi/2$ where $\xi = \sqrt{(\chi_{12} - \chi_{21})^2 + 4\chi_{12}\chi_{21}}$ with the corresponding eigenvectors given as $\psi_\pm = [(\chi_{12} - \chi_{21} \mp \xi)/2\chi_{21} \quad 1]^T$. This system has a spectral degeneracy for $\xi = 0$ which is satisfied when $4\chi_{12}\chi_{21} = -(\chi_{12} - \chi_{21})^2$. The eigenvectors associated with these degenerate eigenvalues are also degenerate and given by $\psi_\pm = [(\chi_{12} - \chi_{21})/2\chi_{21} \quad 1]^T$. Thus, the spectral degeneracies of $H_2$ with $\xi = 0$ are indeed EPs. Scattering properties of the Mie scatterers that leads to EPs can then be obtained from $\xi = 0$. Since $\chi_{12}\chi_{21}$ is a complex number $\chi_{12}\chi_{21} = \epsilon_{121}\epsilon_{112} + \epsilon_{212}\epsilon_{221} + (\epsilon_{121}\epsilon_{212} + \epsilon_{112}\epsilon_{221})\cos\Delta\varphi + j(\epsilon_{112}\epsilon_{221} - \epsilon_{121}\epsilon_{212})\sin\Delta\varphi$ and $(\chi_{12} - \chi_{21})^2$ is a positive real number, we need to set $\text{Im}(\chi_{12}\chi_{21}) = 0$ and $\text{Re}(\chi_{12}\chi_{21}) = -(\chi_{12} - \chi_{21})^2/4$. The former is satisfied for i) $\Delta\varphi = n\pi$ which leads to $(\epsilon_{121}\epsilon_{112} + \epsilon_{212}\epsilon_{221}) - (\epsilon_{121}\epsilon_{212} + \epsilon_{112}\epsilon_{221}) = -(\chi_{12} - \chi_{21})^2/4$, and for ii) $\epsilon_{112}\epsilon_{221} = \epsilon_{121}\epsilon_{212} = \epsilon'$ leading to $\cos\Delta\varphi = -\frac{1}{2}\left(\frac{(\chi_{12}-\chi_{21})^2}{4\epsilon'} + \frac{\epsilon_{112}^2 + \epsilon_{212}^2}{\epsilon_{112}\epsilon_{212}}\right)$. Based on the geometric symmetry of Mie scatterers, we have categorized their combinations in four groups (Supplementary section 2). Complying with our analysis, their $\Delta\varphi$ dependent exemplary eigenvalues exhibit characteristic behavior of DPs and EPs. A combination of symmetric Mie scatterers or identical asymmetric Mie scatterers only leads to DPs in the MRR (**Fig. 1d-e**), while one symmetric one asymmetric Mie scatterer or by two non-identical asymmetric Mie scatterers lead to an EP, if the inter-Mie scatterer distance is properly adjusted (Fig. **1f-g**) (Supplementary Section 2).

Following the theoretical analysis, we fabricated a set of MRRs with embedded symmetric and asymmetric Mie scatterers on a silicon-on-insulator (SOI) substrate (**Fig. 1a-b**). A MRR enhanced directionality is characterized with an add-drop design, where the symmetric ports supports both CW and CCW excitations (**Fig. 1c**). Light input in Port 1 (Port 2) excites the CW (CCW) mode, and the reflected optical powers are collected in Port 4 (Port 3). The high contrast between the reflections from CW and CCW excitations indicate the system operates near an EP (**Fig. 1h**). Using the transmission spectra at the drop-port we have estimated the $Q$ of the fabricated MRR as ~15,000 (**Fig. 1i**).

**Fig. 2** illustrates the design principle of the Mie scatterers enabled EP system. We consider the design is successful when intracavity field becomes chiral (i.e., maximal chirality is achieved when field in CW/CCW direction is zero while the field in the other direction is maximal). In order to achieve this, we used a Smith Chart based 'optical impedance matching' approach where EPs and hence maximal chirality occurs at the intersection of the trajectories of the reflection coefficients of Mie scatterers constructed by tuning the dimensions of the Mie scatterers and the distance between them. At each crossing point, the effective reflections cancel out in the backward or the forward direction. With a symmetric-asymmetric pair of Mie scatterers, the reflection coefficient trajectories (or optical smith chart) change with the width $W$ of the symmetric Mie scatterer and the distance $L$ between the Mie scatterers (**Fig. 2a**). **Fig. 2b** illustrates the case for two asymmetric Mie scatterers. **Figs 2c-d** show the mode-splitting $\xi = \Delta\omega$ as a function of the phase relative phase of $\Delta\varphi$, with $W$ value satisfying impedance matching process in **Figs. 2a-b**. Emergence of EPs, identified by $\xi = |\Delta\omega| = 0$ and the coalesce of both the real and imaginary parts of the reflection $\Delta\omega$, is clearly seen.

One of the challenges in fabricating high-$Q$ resonator systems is mode-splitting arising from intermodal coupling, induced by the distributed nanometer-scale scatters formed during the waveguide etching process. Presence of mode-splitting implies the formation of a standing-wave in the resonator, and suppressing such inter-modal coupling can results in improved $Q$. In order to check the feasibility of our method to overcome this challenge, we studied a resonator system with randomly distributed scattering centers (similar to sidewall roughness which may form due to fabrication imperfections etc.) and a single symmetric Mie scatterer. The sidewall roughness was simulated by randomly distributed Rayleigh scatters whose depth $\sigma$ ranged in the range of 5-50 nm with the correlation length ($L_c$) taken larger than 50 nm [24] (**Fig. S2**). We found that EPs emerge and hence a transition from standing wave to traveling wave is established when the width $W$ and the depth $D$ of the Mie scatterer is tuned to achieve impedance matching between the reflections from the designed Mie scatterer and Rayleigh scatterers (**Figs. 2c** and **2f**). This finding provides a route to suppress inter-modal scattering by embedding judiciously engineered notch geometry in ring resonators. We note here that one can certainly use reflow techniques, although not all the materials can be re-flowed by heating and melting, to reduce the average size of scattering centers but even in such cases of small scatterers inter-modal scattering exists and leads to mode splitting for high-$Q$ resonators [25]. Also, high temperature process beyond 300ºC is prohibited in active silicon devices with doping and metal contacts, as it leads to the device failure.

To experimentally demonstrate the design concept, we fabricated a set of MRRs with same surface roughness levels but each with a symmetric Mie scatterer of different width $W$ chosen in the range 0 - 420 nm. The depth $D$ of the Mie scatterer was chosen as 100 nm to match its reflectance (Fig. S1) with the one from the surface roughness (Fig. S2). **Figs. 3a** and **3b** show the measured transmission and corresponding reflection spectra in the drop and add ports, for this set of MRRs. The geometry dependent parameters are extracted by fitting the coupled mode theory (CMT) model to the measured spectra (solid curves in **Fig. 3a**). Change in the reflection and transmission spectra obtained for Mie scatterers of different $W$ are clearly seen. With notch width away from EP condition, the absence of observable split-mode spectra in the transmission and the presence of non-zero reflection implies that inter-modal coupling due to surface roughness is present in the system. Moreover, the observation that the resonances have deviated from a Lorentzian form and present a broadened form signals the presence of inter-modal coupling. This situation is completely erased for the Mie scatterer with $W = 200$ nm, which is the transmission spectrum presents a Lorentzian line shape and the reflection does not present a resonance and is zero. This corresponds to maximum rectification, which is referred to as the difference between the transmission and reflection spectra, implying that this Mie scatterer completely compensates the inter-modal scattering induced by surface roughness as expected from our theoretical analysis.

Suppression of backscattering is confirmed by the infrared camera captures of the reflected light in port 4 (reflection port) of the MRRs (**Fig. 3c**) which clearly show the presence of a bright spot at port 4 of the MRR without the meta-atom and the dark spot at port 4 of the MRR with the embedded meta-atom (**Fig. 3d**). The transmission (**Fig. 3e**) and reflection spectra (**Fig. 3f**) of MRRs with and without the meta-atom of $W = 200$ nm shows that both MRRs have similar transmission but significantly differing reflection spectra. The measured $Q_t$ increases from 16,300 ($W = 0$ nm) to 21,300 ($W = 200$ nm). Presence of reflection for the MRR without the meta-atom indicates the simultaneous presence of both the CW and CCW light inside the resonator due to scatterer-induced modal coupling although this does not reflect itself as mode-splitting in the transmission spectra. Absence of reflection for the MRR with the Mie scatterer suggests that the system is at an EP and scattering from the Mie scatterer and the backscattering induced by the sidewall scatterers destructively interfere, eliminating the CCW propagating mode. We found that the Mie scatterer with $W = 200$ nm resulted in minimal back-reflection, suggesting a good agreement with our design strategy to approach the EP degeneracy. By comparing to the MRR with surface roughness only, we estimated the reflection parameters of the Mie scatterer as $r_{1b} = 19.6754 e^{j(-1.1381)}$ GHz and $r_{1f} = 17.5753 e^{-j2.5381}$ GHz from the experimental data. Performing model fitting using CMT yields radiation loss $\gamma_o = 13.321$ GHz (i.e., $Q_{in} = 78,360$) and the coupling losses as 39.96 GHz for CW and CCW modes (Supplementary Section 4).

From the measured transmission spectra, we extract the resonance wavelengths at multiple adjacent resonances separated by the free spectral range of ~ 2 nm. The resonance shift versus the Mie scatterer geometry almost the same among the three modes (**Fig. 4a**). Their deviation from the trend (dashed curve) is attributed to the fabrication variation. The resonance shift is proportional to the $W$ dependent diagonal element $\epsilon_{211}$ in equation (1), which comes along with the tuning of off-diagonal elements. We also extracted the off-diagonal element ($\epsilon_{212} e^{-j\Delta\varphi(W)}$) from the reflection spectra. The ratio of the reflection to the transmission at zero-detuning defined as $|\chi_{12}| = \frac{\gamma_t}{2}\sqrt{\frac{R_{1\to 4}}{T_{1\to 3}}}$ where $\gamma_t$ is the total loss rate of the MRR is depicted for these modes in **Fig. 4b**. We see that $|\chi_{12}|$ obtained for these modes are very similar to each other and agree well with the theoretical predicted dependence on $W$ (Supplementary Section 3). These suggest that the system is not sensitive to dispersion. Finally, we experimentally obtained the total quality factor ($Q_t$) from the transmission and found that it linearly decreases with the $\chi_{12}$ extracted from the measured reflection spectra, for the three modes mentioned above. Fig. 4c shows the correlation between transmission and reflection at the resonance frequency, as a function of width $W$. The enhancement of $Q_t$ in the transmission spectra (1.3 times in **Fig. 3e** for the given Rayleigh scattering and 3 times between the maximum and minimum in **Fig. 4c**) is difficult to be captured in previous studies which used nanotips for EP but it can be systematically studied in our the integrated photonic MRR. We note that this robustness enabled by the integrated photonic platform is highly desirable in EP studies [26-27].

**Discussion**

In conclusion, we have shown that non-Hermiticity in optical resonators can be engineered by designing and lithographically patterning Mie scatterers into the resonator. The amplitude and phase of the Mie scatterer's reflectivity can be deterministically controlled by its *D* and *W*, respectively [22]. The Mie scatterers embedded in MRRs offer a unique low loss approach because the perturbed propagation vector still satisfies the total internal reflection for the guided modes in the resonator. This has not only allowed us to bring a system to an EP but also allowed us to go beyond the fabrication induced Rayleigh scattering integrated photonic high-*Q* MRR and micro-disk resonators [32-33]. Detailed performance matrix comparison of this predefined EP approach and the ones in literature is summarized in Supplementary Section 5.

Non-Hermicity is ubiquitous in integrated photonic resonators but not well formulated, neither of their chirality [28-36]. In all the devices, the random waveguide sidewall roughness introduces different off-diagonal elements (Fig. S2) in the Hamiltonian and leads to unexpected mode-splitting in MRR [28-29, 34-35]. Counterintuitively, here we show that additional 'defect' of the Mie scatterer can further enhances *Q* (~$10^5$) by suppressing the coherent backscattering. This post-tuning free chiral integrated photonic system opens many opportunities towards EP-enhanced sensors, optomechanical and parametric nonlinear interactions [36-37], and lasing. It also helps to build photonic structures with asymmetric reflection profiles and backscattering-free resonators for various applications such as frequency combs, solitons, sensing, and other nonlinear optical processes such as photon blockade, and regenerative oscillators. The lithographically defined platform enables deterministic control of light propagation, toward the exploration of EP physics coupled with parametric and free-carrier dynamics in silicon MRRs.

Beyond opening a new direction for chirality silicon photonics, the combined theoretical analysis and experimental results that presented in this work also advances the knowledge of EP [38]. First, it reveals the critical role of spatial asymmetry of the nanotip and Mie scatterers for bringing the system towards EP. Second, scatter geometry-controlled pathway of driving the non-Hermitian system towards and away from an EP is illustrated in details. Third, the mechanically stable system allows reliable comparison between transmission and reflection spectra for the perturbed microresonator, which reveals the nanotip/scatter's contribution to the diagonal terms. Forth, enhancement of the empirical *Q* extracted from the transmission spectra are demonstrated for the first time.

**Materials and methods**

**Sample fabrication.** The MRRs are fabricated on a SOI substrate from Soitec, having a 250 nm silicon layer and 3 µm buried oxide layer. The substrates initially undergo pre-cleaning in organic solvents followed by Nanostrip at 80°C to remove any particulate matter or contaminants on the surface. The structures are patterned in AR-P-CSAR 6200.09 positive E-beam resist (thickness: ~ 275 nm) by Vistec EBPG5200 electron beam lithography system with 100 kV acceleration voltage. A base dosage is adjusted to be ~120 $\mu c^2/cm^2$ for the multi-pass writing to minimize the field stitching error, followed by a proper proximity effect correction. During the electron beam lithography, a ring waveguide is broken down into a polygon path, resulting in different number of actual exposure dots around the path. The pattern is developed using AR-600-546 developer solution. Our chip is then etched by inductively coupled plasma (ICP) Etching tool using Fluorine-based chemistry, in which Sulfur hexafluoride ($SF_6$) gas is used to mainly etch the silicon while Octafluorocyclobutane ($C_4F_8$) gas is carefully adjusted with respect to $SF_6$ gas for the passivation during etching, preventing a thin E-beam resist erosion ($C_4F_8$:$SF_6$=1:1). An intermittent plasma asher step is performed to remove any residual resist. A 0.3 $\mu$m thick silicon dioxide is deposited by plasma enhanced chemical vapor deposition (PECVD) at 300°C and 1.4 Torr pressure using silane ($SiH_4$) and Nitrous oxide ($N_2O$) as precursor gases.

**Optical measurements.** A tunable laser in telecommunication C band sends the TE polarized light to the on-chip grating coupler through a polarization controller. The output power is monitored by a Newport InGaAs photodiode (818-IG-L-FC/DB) and an optical power meter (1830-R-GPIB).

**Numerical simulations.** A three-dimensional finite-difference-time-domain (3D FDTD) method simulates the optical field distribution, transmission, and reflection coefficients. The conformal mesh with spatial resolution less than 1/10 of the local feature size is applied for FDTD simulation.

**Availability of data and materials:** The datasets used and/or analyzed during the current study are available from the corresponding author on reasonable request.

**Competing interests:** All other authors declare they have no competing interests.

**Funding:** This work is supported by Defense Advanced Research Projects Agency (N660012114034) and Air Force Office of Scientific Research (AFOSR) Multi-University Research Initiative (FA9550-21-1-0202). SKÖ acknowledges AFOSR (FA9550-18-1-0235). The device fabrication is partially supported by AFOSR (FA9550-18-1-0300).

**Author contributions:**
H. L. and T. G. conceived the idea. H. L., S. K. O. and T. G. developed the experimental plan. H. L. designed and fabricated the devices. H. L. and F. W. conducted the device characterizations. H. L. and A. K. performed the detailed modeling and results analysis, supervised by S. K. H. L. and

T. G. wrote the manuscript with inputs from the other coauthors.

**Acknowledgments**

The devices are fabricated at the University of Delaware Nanofabrication Facility with assistance from Dr. Kevin Lister.

# Figures

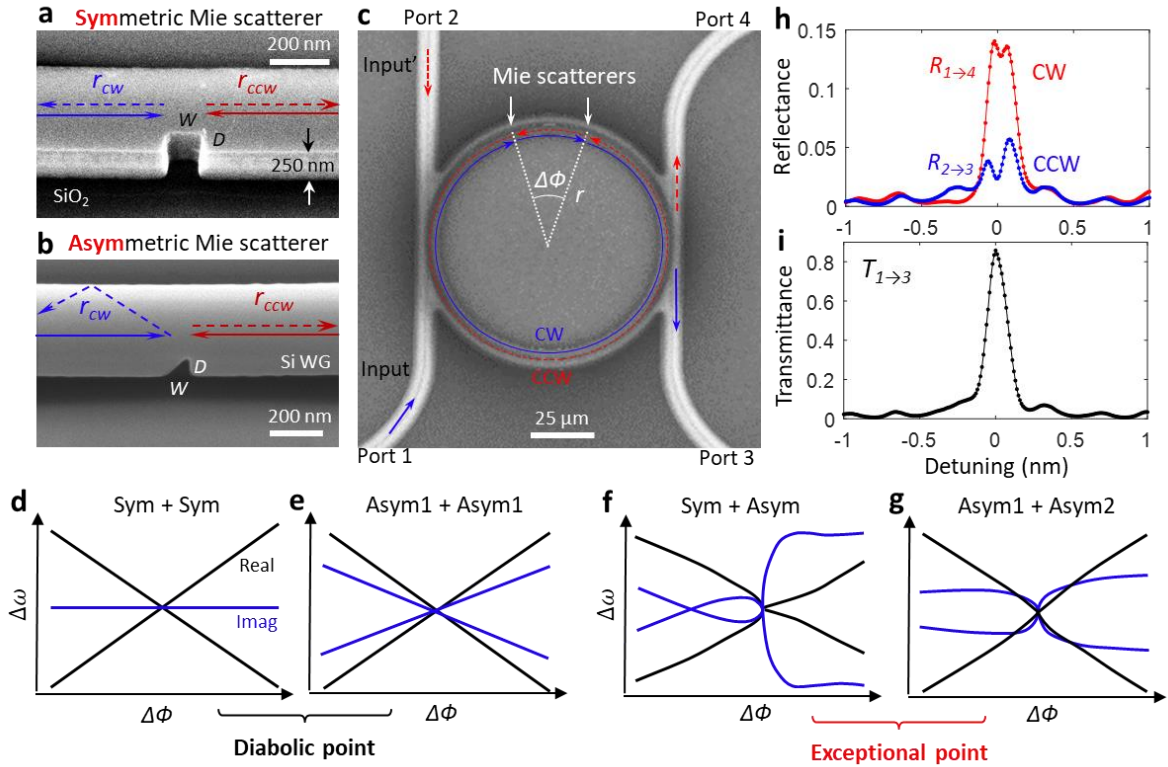

**Figure 1 | Exceptional points enabled by a pair of Mie scatterers embedded in channel waveguides and resonators.** (a) Scanning electron microscope (SEM) image of a channel waveguide (WG) with a lithographically defined rectangular-shaped symmetric Mie-scatterer on SOI substrate. (b) SEM image of a typical asymmetric Mie scatterer with right triangle shape. (c) A microring resonator (MRR) based add-drop filter with Mie scatterers. The Mie scatterers (a,b) defined on the ring perimeter control the non-Hermiticity and spectral degeneracy of the MRR state. Depths ($D$) and widths ($W$) are first chosen such that the two Mie scatterers have the same reflectance (Fig. S1). Then the optical path difference $\varDelta\varPhi$ between the Mie scatterer pair is tuned to bring the system to an EP or DP. (d, e) A system with a pair of identical symmetric or asymmetric Mie scatterers leads only to a diabolic point (DP) spectral degeneracy. (f, g) A combination of one asymmetric and one symmetric Mie scatterer or two non-identical asymmetric Mie scatterers may be tuned to create EPs. The real and imaginary parts of the eigenmodes are plotted in black and blue, respectively. (h) Measured asymmetric reflection spectra (port 1 to 4 and port 2 to 3) and (i) transmission spectrum (port 1 to 3) of the add-drop filter with optimized geometric parameters creating an EP. The asymmetry in the reflection spectra for clockwise (CW) and counterclockwise (CCW) inputs in (h) suggests that the system is at an EP.

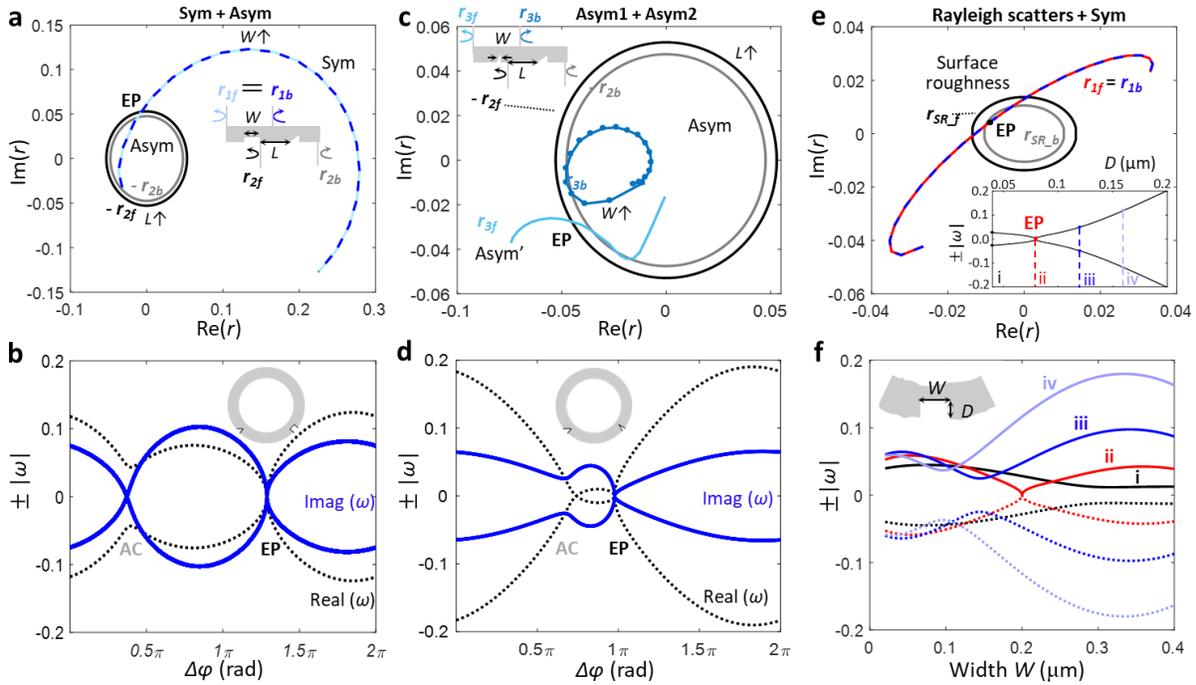

**Figure 2 | Optical impedance matching towards EP.** (a) Complex reflectivity of a Mie scatterers embedded channel waveguide with forward ($r_f$) and backward ($r_b$) excitation. The symmetric rectangular shaped Mie scatterers with dimensions $D = 200$ nm, $W = 20 \sim 400$ nm have the same $r_f$ (light blue solid curve) and $r_b$ (blue dashed curve). An asymmetric (right triangle) Mie scatterer with dimension $D = 200$ nm and $W = 140$ nm has different $r_f$ (black curve) and $r_b$ (gray). Distance $L$ between the Mie scatterers is varied in the range $L = 0 \sim \lambda$. (b) Asymmetric reflectivity of a channel waveguide with two asymmetric (i.e., right triangle) Mie scatterers. Blue curves: $D_{Asym'} = 160$ nm, $W_{Asym'} = 20$ to $400$ nm. Black and grey curves: $D_{Asym} = 200$ nm, $W_{Asym} = 140$ nm. The optical path length between the input point and the scatter varies from 0 to $\lambda$. (c,d), The real (black dashed curve) and imaginary (blue dots) parts of the eigenvalues versus optical path length $\Delta\Phi$ (c) between a symmetric and an asymmetric Mie scatterers, and (d) between two nonidentical asymmetric Mie scatterers. $W$ of the symmetric Mie scatterer in (c) is taken as 80 nm, and that of the asymmetric Mie scatterer in (d) is taken as 240 nm. The dimensions of the Mie scatterers used in (c) and (d) are chosen from the intersection points of reflectivity plots in (a) and (b). EP: exceptional point; AC: avoided crossing. (e) Complex reflectivity of a channel waveguide with an embedded symmetric Mie scatterer (red: forward scattering; blue: backward scattering) and that of a channel waveguide with surface roughness (black: forward scattering; gray: backward scattering). The Mie scatterer depth $D$ is fixed at 80 nm, and its width $W$ is varied from 20 nm to 400 nm. The
14

reflectivity trajectory of the Mie-scatterer overlaps with the reflectivity of the channel waveguide with sidewall surface roughness ($L_c = 100$ nm, $\sigma = 10\ nm$) at $W$=180 nm. Inset: Magnitude of the complex frequency splitting $\pm|\omega|$ versus $D$ of the Mie-scatterer with $W$=180 nm defined on the perimeter of the MRR. EP emerges at $D$ value where $\pm|\omega| = 0$. (f) $\pm|\omega|$ versus W of the Mie-scatterer with $D$ equals to 30 nm (black), 80 nm (red), 120 nm (blue) and 160 nm (light blue).

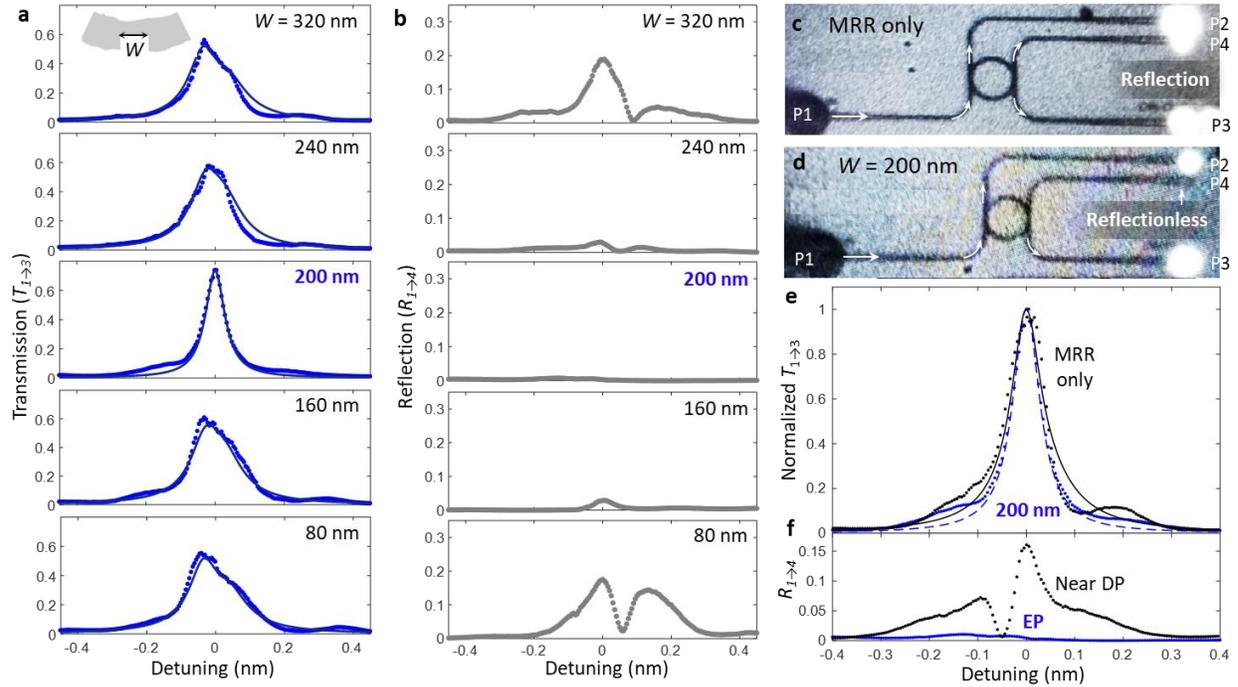

**Figure 3 | Experimental demonstration of suppressed reflection and improved quality factor in MRRs by engineering geometry of the Mie scatterer.** (a) Transmission ($T_{1\to3}$) and (b) reflection spectra ($R_{1\to4}$) of the MRR in add-drop configuration as shown in Fig. 1 for a Mie scatterer with $D = 100$ nm and $W = 80$ nm, 160 nm, 200 nm, 240 nm, and 320 nm (top to bottom). The devices with different $W$ are fabricated in the same run to make sure that they are affected in the same way during the fabrication (e.g., similar fabrication imperfections and sidewall roughness). $R_{1\to4}$ becomes zero, suggesting the emergence of an EP, at $W = 200$ nm in good agreement with the design process in Fig. 2. (c) Infrared camera captured top view of a ring resonator without and (d) with the Mie scatterer ($W = 200$nm, $D = 100$nm). A single mode fiber launches light through the grating coupler connected to port 1. The ring resonator without the Mie scatterer experiences intermodal scattering due to sidewall roughness which results in strong reflection signal at port 4 (reflection port), which is suppressed in d. (e) $T_{1\to3}$ and (f) $R_{1\to4}$ of the MRRs shown in the c (black) and d (blue), respectively. Dots: experimental data. Curves: CMT fits.

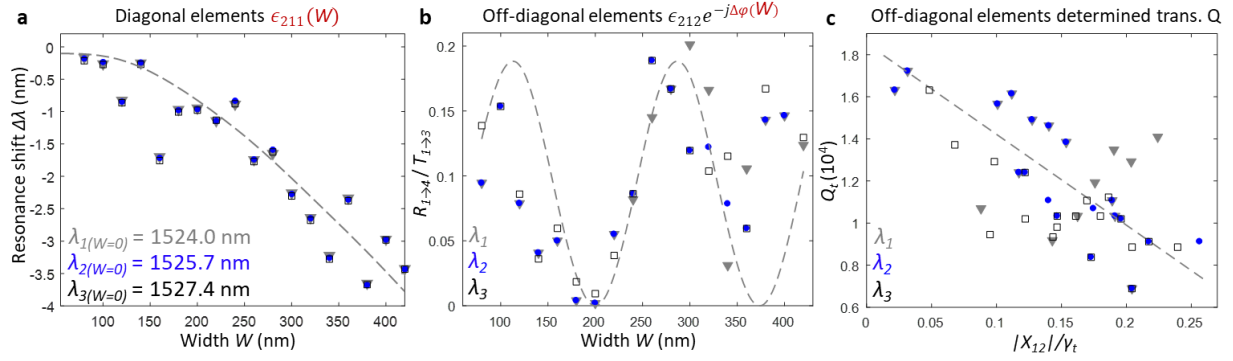

**Figure 4 | Measured dependence of diagonal and off-diagonal elements in the non-Hermitian Hamiltonian (equation 1) on the Mie scatterer width.** (a) Resonance wavelength detuning versus the notch width $W$ for three adjacent modes marked as $\lambda_{1-3}$, with the incremental step of $\Delta W = 20$ nm. The resonance wavelength of the modes is between 1520-1530nm. (b) Comparison of measured and modeled (dashed curve) on-resonance transmission and reflection versus designed Mie scatterer width $W$. The grey curve is CMT mode prediction. (c) Total quality factor extracted from the transmission spectra ($T_{1\rightarrow 3}$) versus normalized forward scattering coefficient $|X_{12}|/\gamma_t$ (extracted from the normalized reflection $R_{1\rightarrow 4} / T_{1\rightarrow 3}$). The grey triangles, blue dots and black squares are experimental data for the three adjacent modes separated by 2 nm free spectral range. The dashed grey curve and line in a and c are eye guiders. The grey curve in b is predicted by CMT model.

# Supplementary Materials for

## Chiral exceptional point and coherent suppression of backscattering in silicon microring with low loss Mie scatterer

Hwaseob Lee, Ali Kecebas, Feifan Wang, Lorry Chang, Sahin K. Özdemir*, Tingyi Gu[*]

*Corresponding author. Email: tingyigu@udel.edu, sko9@psu.edu

Supplementary Text
Sec 1: Directional reflection coefficients in notch embedded single mode waveguide
Sec 2: Mie scatterer defined non-Hermicity in MRRs
Sec 3: Empirical model for extracting the experimental reflection parameters
Sec 4: Interpretation of experimental data
Sec 5: Performance comparison
Figs. S1 to S5
References (S1 to S13)
Table S1

**Supplementary Section 1.** Directional reflection coefficients in notch embedded single mode waveguide

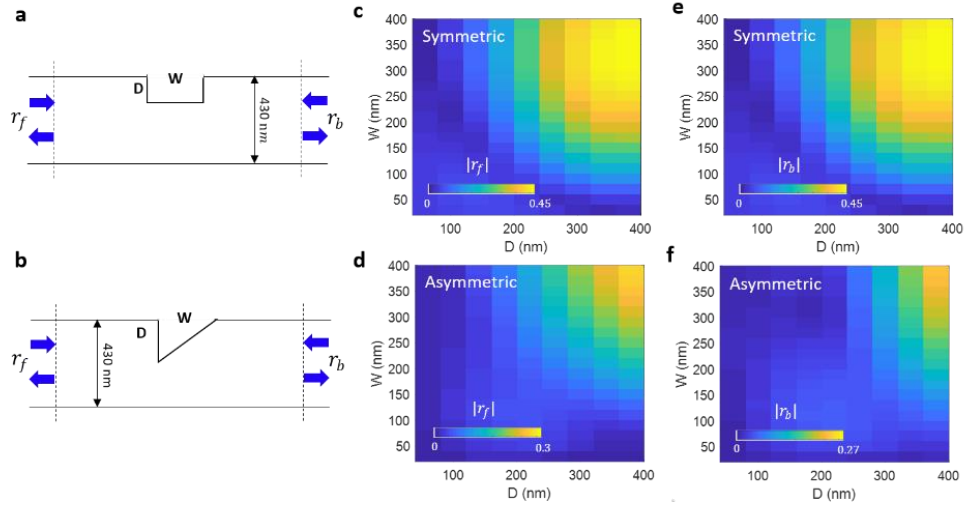

**Fig. S1. Width (W) and Depth (D) determined reflectance for symmetric and asymmetric Mie scatterers.** (a) Schematics of the rectangular notch reflector as a typical symmetric Mie scatterer, and (b) right-angled triangular asymmetric reflector. The reflection coefficients with forward excitation and backward excitations are marked as $r_f$ and $r_b$ respectively. (c) Amplitude of the forward reflectivity ($r_f$) versus the notch depth $D$ for the symmetric and (d) asymmetric notches. (e) Amplitude of the backward reflectivity ($r_b$) for the symmetric reflector and (f) asymmetric notch.

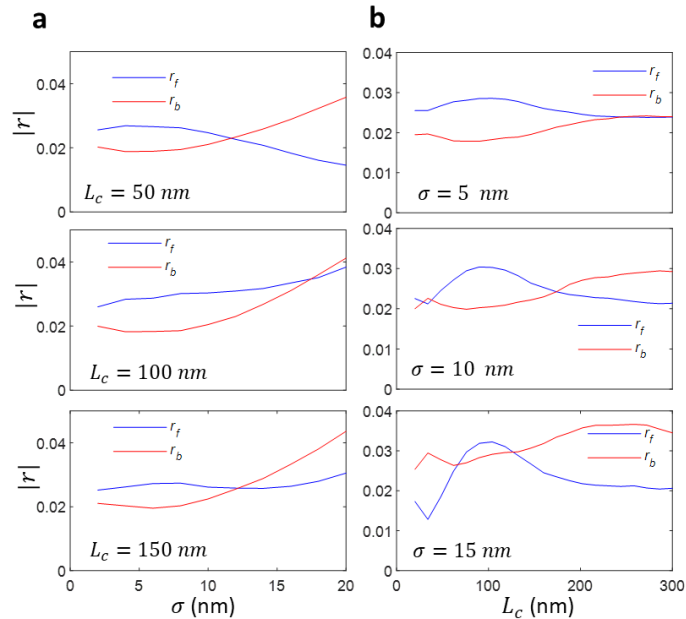

**Fig. S2. Distributed Rayleigh scatterer induced asymmetric reflection.** (a) The correlation length ($L_c$) dependent asymmetric reflectivity amplitude fixed roughness ($\sigma = 10\ nm$). (b) The roughness magnitude ($\sigma$) dependent asymmetric backscattering with fixed correlation length ($L_c = 100\ nm$).

Carried out by the three-dimensional full field simulation, Fig. S1 shows the embedded Mie scatterer geometry defined symmetric (Fig. S1a,c,e) and asymmetric reflectance (Fig. S1b,d,f). In three-dimensional Finite-Difference-Time-Domain simulation [S1], the conformal mesh with a spatial resolution (in $\hat{x}$ and $\hat{y}$ direction) less than 1/10 of the Mie scatterer dimension is applied. For the vertical direction ($\hat{z}$), the spatial resolution of the mesh is fixed as 1/5 of the simulated silicon waveguide thickness. The simulated waveguide structure is extended completely through PML region for the stable and accurate results. The wavelength of the incident light is in the telecommunication C band (1531 nm). The channel waveguide with width of 430 nm width is defined on 250 nm thick silicon-on-insulator substrate.

With the same methodologies, we simulate the directional reflectivity from the waveguides with fabrication-related sidewall roughness (E-beam lithography + F-ICP etching). The roughness distribution is asymmetric in parallel and symmetric in vertical direction to the direction of wave propagation as shown in Fig. S2 [S2]. We simulate the random geometry with correlation function ($\sigma$: roughness magnitude, $L_c$: correlation length) [S3]. Fig. S2 confirms the random sidewall roughness generates mostly asymmetric reflectivity [$r_f \neq r_b$].

Note $r_{f/b}$ is presented as $r_{CW/CCW}$ for the Mie scatterer embedded in the microring resonator (MRR).

**Supplementary Section 2: Mie scatterer defined non-Hermicity in MRRs**

A MRR without the Mie scatterers is described by the Hamiltonian $H_o = \begin{pmatrix} \Omega_0 & 0 \\ 0 & \Omega_0 \end{pmatrix}$, where the diagonal elements $\Omega_0 = \omega_o - j\gamma_t/2$ represent the complex energies of the degenerate CW and CCW modes, $\gamma_t$ representing all losses including both the intrinsic losses ($\gamma_0$, including material, bending, ring geometry dependent loss) and the resonator-waveguide coupling losses rates ($\gamma_c$). This system supports a DP with the degenerate eigenfrequencies $\omega_{1,2} = \omega_o - j\gamma_t/2$ and orthogonal eigenvectors $\psi_1 = [1 \quad 0]^T$ and $\psi_2 = [0 \quad 1]^T$ corresponding to CW and CCW modes.

**2.1 MRR with single Mie scatterer**

A first meta-atom introduced into the mode volume of the resonator will not only couple the CW and CCW modes through intermodal scattering, but it will also induce a complex frequency shift. Then the Hamiltonian describing the resonator with the first embedded meta-atom is given by

$$H_1 = \begin{pmatrix} \Omega_0 + \epsilon_{111} & \epsilon_{112} \\ \epsilon_{121} & \Omega_0 + \epsilon_{122} \end{pmatrix} \tag{S-1}$$

whose complex-valued elements $\epsilon_{1km}$ describes the coupling strength of the particle-induced scattering into the same ($k = m$) or the counterpropagating ($k \neq m$) mode. The diagonal elements $\epsilon_{111}$ and $\epsilon_{122}$ determine the frequency shift and the additional dissipation induced by the Mie scatterer to the CW and CCW modes whereas the off-diagonal elements $\epsilon_{112}$ and $\epsilon_{121}$ determine the coupling between them. The eigenfrequencies of this system are found as $\omega_\pm = \Omega_0 + (\epsilon_{111} + \epsilon_{122})/2 \pm \xi/2$ where $\xi = \sqrt{(\epsilon_{111} - \epsilon_{122})^2 + 4\epsilon_{112}\epsilon_{121}}$ implying a splitting of $\omega_+ - \omega_- = \xi$. A nice property of Mie scatterers embedded in MRRs is that they induce negligible loss as the reflected light still complies with the total internal reflection on the waveguide sidewalls, and couples into the guided mode in the resonator. Without loss of generality, $\epsilon_{1km}$ can be treated as real valued ($\epsilon_{1km} = \epsilon_{1kmr} + i\epsilon_{1kmi}$ where $|\epsilon_{1kmi}| \ll |\epsilon_{1kmr}|$) [22].

### 2.1.1 Asymmetric Mie scatterer

It is easy to see that there exists a spectral degeneracy of $\omega_\pm = \Omega_0 + \epsilon$ when $\epsilon_{111} = \epsilon_{122} = \epsilon$ and $\epsilon_{112}\epsilon_{121} = 0$. Under these settings, the Hamiltonian describing the system becomes $H_{1a} = \begin{pmatrix} \Omega_0 + \epsilon & \epsilon_{112} \\ 0 & \Omega_0 + \epsilon \end{pmatrix}$ or $H_{1b} = \begin{pmatrix} \Omega_0 + \epsilon & 0 \\ \epsilon_{112} & \Omega_0 + \epsilon \end{pmatrix}$ which describes the cases in which there is no backscattering into the CW (or CCW) mode even though there is coherent backscattering into the CCW (or CW). The eigenvectors described by these Hamiltonians are also degenerate and are given as $\psi_\pm \propto [1 \quad 0]^T$. Thus, this spectral degeneracy is an EP.

### 2.1.2 Symmetric Mie scatterer

For a symmetric Mie scatterer (e.g., rectangular notch in Fig 1a), on the other hand, we have $\epsilon_{111} = \epsilon_{122} = \epsilon_1$ and $\epsilon_{112} = \epsilon_{121} = \epsilon_1'$ which results in $\omega_\pm = \Omega_0 + \epsilon_1 \pm \epsilon_1'$ with the corresponding orthogonal eigenvectors $\psi_\pm = [1 \quad \pm 1]^T$. Thus, the complex eigenfrequencies differ by $2\epsilon_1'$, which shows itself as two spectrally different resonance peaks in the reflection spectra of the resonator coupled waveguide system. This splitting has formed the basis for mode-splitting based particle detection systems [1]. Thus, for a single symmetric scatterer, the system does not have an EP. This is true even for the case $\epsilon_{111} = \epsilon_{122} = \epsilon_{112} = \epsilon_{121}$.

### 2.2 MRR with a Mie scatterer pair

### 2.2.1 General case of two Mie scatterers (Asym1 + Asym2)

A MRR with two embedded Mie scatterers will provide additional degrees of freedom to tune the system towards or away from its spectral degeneracies and help to counterbalance the fabrication imperfections. One can write a general Hamiltonian for such a system as

$$H_2 = \begin{pmatrix} \Omega_0 + \epsilon_{111} + \epsilon_{211} & \epsilon_{112} + \epsilon_{212}e^{-j\Delta\varphi} \\ \epsilon_{121} + \epsilon_{221}e^{j\Delta\varphi} & \Omega_0 + \epsilon_{122} + \epsilon_{222} \end{pmatrix} = \begin{pmatrix} \Omega_0 + \chi_1 & \chi_{12} \\ \chi_{21} & \Omega_0 + \chi_2 \end{pmatrix} \quad \text{(S-2)}$$

The eigenvalues of $H_2$ are $\omega_\pm = \Omega_0 + (\chi_1 + \chi_2)/2 \pm \xi/2$ where $\xi = \sqrt{(\chi_1 - \chi_2)^2 + 4\chi_{12}\chi_{21}}$ with the corresponding eigenvectors given as $\psi_\pm = [(\chi_1 - \chi_2 \mp \xi)/2\chi_{21} \quad 1]^T$. This system has a spectral degeneracy for $\xi = 0$ which is satisfied when $4\chi_{12}\chi_{21} = -(\chi_1 - \chi_2)^2$. The eigenvectors associated with these degenerate eigenvalues are also degenerate and given by $\psi_\pm = [(\chi_1 - \chi_2)/2\chi_{21} \quad 1]^T$. Thus, the spectral degeneracies of $H_2$ with $\xi = 0$ are indeed EPs. Scattering properties of the Mie scatterers that leads to EPs can then be obtained from $\xi = 0$. Since $\chi_{12}\chi_{21}$ is a complex number $\chi_{12}\chi_{21} = \epsilon_{121}\epsilon_{112} + \epsilon_{212}\epsilon_{221} + (\epsilon_{121}\epsilon_{212} + \epsilon_{112}\epsilon_{221})\cos\Delta\varphi + j(\epsilon_{112}\epsilon_{221} - \epsilon_{121}\epsilon_{212})\sin\Delta\varphi$ and $(\chi_1 - \chi_2)^2$ is a positive real number, we need to set $\text{Im}(\chi_{12}\chi_{21}) = 0$ and $\text{Re}(\chi_{12}\chi_{21}) = -(\chi_1 - \chi_2)^2/4$. The former is satisfied for i) $\Delta\varphi = n\pi$ which leads to $(\epsilon_{121}\epsilon_{112} + \epsilon_{212}\epsilon_{221}) - (\epsilon_{121}\epsilon_{212} + \epsilon_{112}\epsilon_{221}) = -(\chi_1 - \chi_2)^2/4$, and for ii) $\epsilon_{112}\epsilon_{221} = \epsilon_{121}\epsilon_{212} = \epsilon'$ leading to $\cos\Delta\varphi = -\frac{1}{2}\left(\frac{(\chi_1-\chi_2)^2}{4\epsilon'} + \frac{\epsilon_{112}^2 + \epsilon_{212}^2}{\epsilon_{112}\epsilon_{212}}\right)$. Clearly, an experimental ability to tune the optical phase delay between the Mie scatterers (i.e., the phase $\Delta\varphi$) may help to move the system closer to or away from the spectral degeneracy.

### 2.2.2 One symmetric and one asymmetric Mie scatterer (Sym + Asym)

The analysis presented above implies that having a symmetric and an asymmetric Mie scatterer will provide the necessary flexibility to move the system towards or away from an EP and study the associated processes. The Hamiltonian for such a system with one symmetric Mie scatterer (e.g., the first Mie scatterer with $\epsilon_{111} = \epsilon_{122} = \epsilon_1$ and $\epsilon_{112} = \epsilon_{121} = \epsilon_1'$) and one asymmetric Mie scatterer with scattering properties defined, without loss of generality, as $\epsilon_{211} = \epsilon_{222} = \epsilon_2$, $\epsilon_{212} = \epsilon_2'$, and $\epsilon_{221} = \epsilon_2''$ becomes

$$H_{2h} = \begin{pmatrix} \Omega_0 + \epsilon_1 + \epsilon_2 & \epsilon_1' + \epsilon_2' e^{-j\Delta\varphi} \\ \epsilon_1' + \epsilon_2'' e^{j\Delta\varphi} & \Omega_0 + \epsilon_1 + \epsilon_2 \end{pmatrix} = \begin{pmatrix} \Omega_0 + \chi_1' & \chi_{12}' \\ \chi_{21}' & \Omega_0 + \chi_2' \end{pmatrix} \quad \text{(S-3)}$$

whose eigenvalues are found $\omega_\pm = \Omega_0 + \epsilon_1 + \epsilon_2 \pm \xi$ where $\xi = \sqrt{\epsilon_1'^2 + \epsilon_2'\epsilon_2'' + (\epsilon_2' + \epsilon_2'')\cos\Delta\varphi - j(\epsilon_2' - \epsilon_2'')\sin\Delta\varphi}$. This system has a spectral degeneracy when $\xi = 0$ with the associated eigenvectors found as $\psi_\pm = [0 \ 1]^T$ which is an EP. For this EP to emerge, scattering properties of the Mie scatterers should satisfy $\epsilon_2' - \epsilon_2'' = 0$ or $\sin\Delta\varphi = 0$. For the former, the real part becomes $\epsilon_1'^2 + \epsilon_2'^2 + 2\epsilon_2'\cos\Delta\varphi$ which may be made zero by tuning the Mie scatterer properties and the phase to satisfy $\cos\Delta\varphi = -(\epsilon_1^2 + \epsilon_2'^2)/2\epsilon_2'$. For the latter, we have $\sin\Delta\varphi = 0$ and thus $\epsilon_1'^2 = \epsilon_2' + \epsilon_2'' - \epsilon_2'\epsilon_2''$.

### 2.2.3 Different symmetric Mie scatterers (Sym1 + Sym2)

It is easy to see that for two-symmetric Mie scatterers with $\epsilon_{111} = \epsilon_{122} = \epsilon_1$, $\epsilon_{211} = \epsilon_{222} = \epsilon_2$, $\epsilon_{112} = \epsilon_{121} = \epsilon_1'$ and $\epsilon_{212} = \epsilon_{221} = \epsilon_2'$ ), the Hamiltonian becomes

$$H_{2s} = \begin{pmatrix} \Omega_0 + \epsilon_1 + \epsilon_2 & \epsilon_1' + \epsilon_2' e^{-j\Delta\varphi} \\ \epsilon_1' + \epsilon_2' e^{j\Delta\varphi} & \Omega_0 + \epsilon_1 + \epsilon_2 \end{pmatrix} \quad \text{(S-4)}$$

whose eigenvalues are $\omega_{+/-} = \Omega_0 + \epsilon_1 + \epsilon_2 \pm \xi$ with $\xi = \sqrt{\epsilon_1'^2 + \epsilon_2'^2 + 2\epsilon_1'\epsilon_2'\cos\Delta\varphi}$. The degeneracy will be observed when $\xi = 0$, implying $\cos\Delta\varphi = -(\epsilon_1'^2 + \epsilon_2'^2)/2\epsilon_1'\epsilon_2'$. The eigenvectors associated with this degeneracy are $\psi_\pm \propto [0 \ 1]^T$ which also implies that this degeneracy is an EP degeneracy. The opposite signs of the phases allow one to make one of the off-diagonal terms zero while keeping the other non-zero. If the symmetric Mie scatterers are identical (i.e., $\epsilon_{111} = \epsilon_{122} = \epsilon_{211} = \epsilon_{222} = \epsilon_1$ and $\epsilon_{112} = \epsilon_{212} = \epsilon_{121} = \epsilon_{221} = \epsilon_1'$), it is easy to show that eigenvalues are $\omega_\pm = \Omega_0 + 2\epsilon_1 \pm 2\epsilon_1' \cos\Delta\varphi$ with the corresponding eigenvectors found as $\psi_\pm \propto [1 \ \pm 1]^T$ which is independent of $\Delta\varphi$. Therefore, although the eigenvalues become degenerate for $\Delta\varphi = (2n + 1)\pi/2$, the eigenvectors remain orthogonal. Thus, we conclude that a system with two identical symmetric Mie scatterers exhibits only a DP (Fig. 1f). Note that by tuning the angle $\Delta\varphi$, one can vary the splitting between zero and the maximum value of $4\epsilon_1'$. Therefore, to obtain an EP, one needs to design the Mie scatterers to be sufficiently different to provide the asymmetry in the off-diagonal terms.

### 2.2.4 Identical symmetric Mie scatterers (Asym1 + Asym1)

One naturally asks what will happen if two identical asymmetric Mie scatterers with $\epsilon_{111} = \epsilon_{122} = \epsilon_{211} = \epsilon_{222} = \epsilon$, $\epsilon_{112} = \epsilon_{212} = \epsilon'$, and $\epsilon_{121} = \epsilon_{221} = \epsilon''$. Hamiltonian in this case becomes:

$$H_{2s} = \begin{pmatrix} \Omega_0 + 2\epsilon & \epsilon'(1 + e^{-j\Delta\varphi}) \\ \epsilon''(1 + e^{j\Delta\varphi}) & \Omega_0 + 2\epsilon \end{pmatrix} \quad \text{(S-5)}$$

whose eigenvalues are $\omega_\pm = \Omega_0 + 2\epsilon \pm \xi$ with $\xi = \sqrt{2\epsilon'\epsilon''(1 + \cos\Delta\varphi)}$. The degeneracy will be observed when $\xi = 0$, implying $\epsilon'\epsilon'' = 0$ or $\cos\Delta\varphi = (n + 1)\pi$. When $\epsilon' = \epsilon'' = 0$ or $\cos\Delta\varphi = (n + 1)\pi$, the degeneracy is a DP (i.e., both off-diagonal elements are zero), when either $\epsilon'$ or $\epsilon''$ is zero, the degeneracy is an EP (i.e., unidirectional coupling between CW and CCW modes) but as we discussed above creating such Mie scatterers are challenging. Thus, having the phase $\Delta\varphi$ as the only tunable parameter does not allow creating an EP when the Mie scatterers are identical and asymmetric. It is easy to extend the above analysis to the case of two non-identical asymmetric Mie scatterers showing that the spectral degeneracy is an EP and phase can be tuned $\Delta\varphi$ to arrive at this EP.

### 2.2.5 Summary for inter-scatter phase defined EP

The exemplary results for the four cases discussed in sections 2.2.1-2.2.4, at which variation of eigenvalues with respect to $\Delta\varphi$, are plotted in Fig. S3.

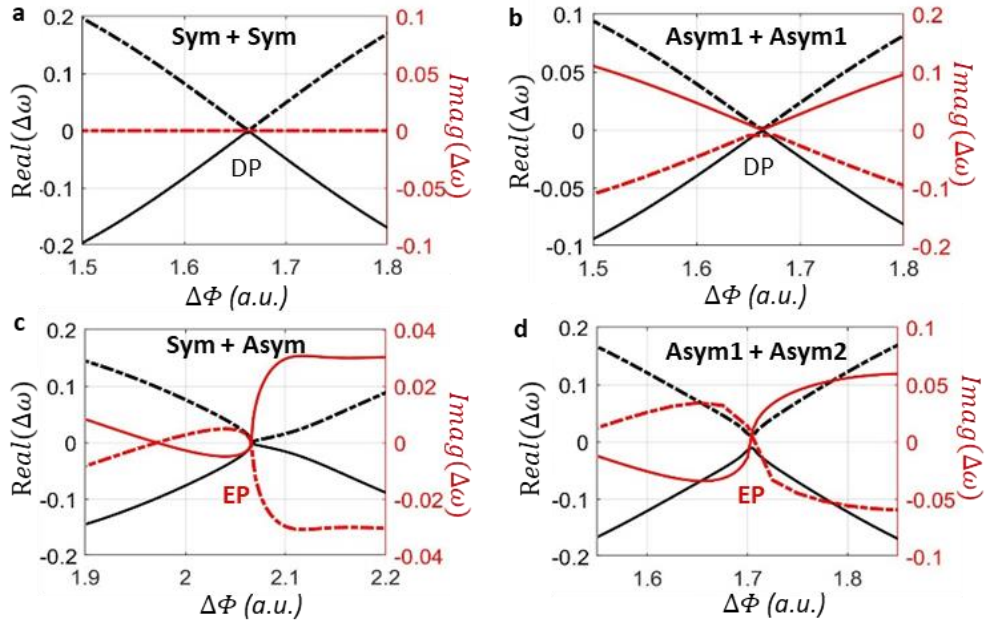

**Fig. S3. Mie scatterer combinations and resulting non-Hermitian states.** Splitting in the real and imaginary parts of the eigenfrequency as a function of the inter-Mie scatterers distance with (a) two symmetric scatterers, (b) identical asymmetric scatterers, (c) one symmetric and one asymmetric scatterer and (d) different asymmetric scatterers.

### 2.2.6 Additional note on the scatter geometry defined EP

Variation of eigenvalues with respect to $\Delta\varphi$ and the effect of introducing asymmetry in one of the scatterers is demonstrated in Fig. S4. When the scatterers are identical and symmetric, eigenvalues of the system coalesce, and a DP degeneracy emerges when $\Delta\varphi = \pi/2$. When one of the scatterers is asymmetric with $\epsilon_2'' = \epsilon_2' + \Delta\epsilon$, an EP degeneracy emerges when $\Delta\varphi = \pi$ and $\epsilon_1'^2 = \epsilon_2' + \epsilon_2'' - \epsilon_2'\epsilon_2''$ conditions are satisfied simultaneously.

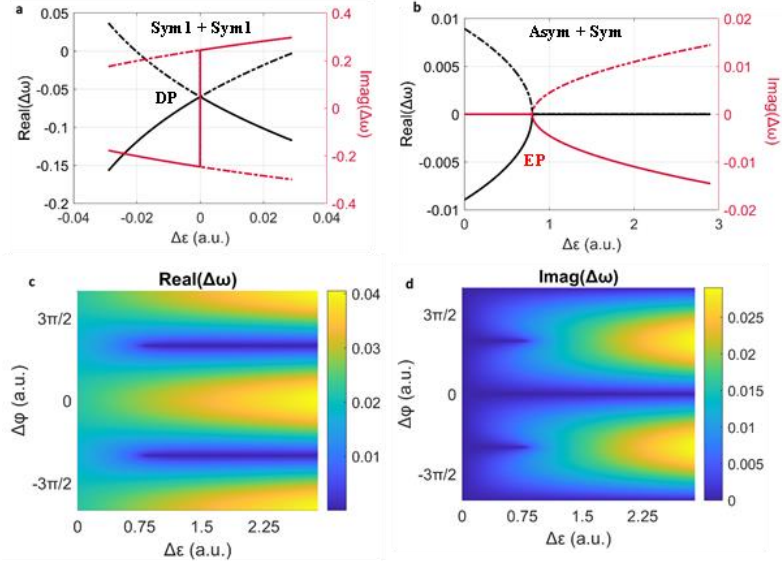

**Fig. S4. Mie scatterer geometry defined non-Hermitian states.** (a) Two identical scatterers and (b) two non-identical scatterers.

## Supplementary Section 3: Empirical model for extracting the experimental parameters

Under slow varying approximation and two mode approximation, the field amplitude for clockwise (CW) and counterclockwise (CCW) propagating mode ($a_{cw/ccw}$) in an MRR (Fig. S5) can be described by a coupled harmonic oscillator model [S4-S7]:

$$\frac{da_{cw}}{dt} = i\Delta\omega a_{cw} - \frac{\gamma_{t\_cw}}{2} a_{cw} - i\chi_{12} a_{ccw} + \sqrt{\gamma_{c1}} s_1 \tag{S-6-1}$$

$$\frac{da_{ccw}}{dt} = i\Delta\omega a_{ccw} - \frac{\gamma_{t\_ccw}}{2} a_{ccw} - i\chi_{21} a_{cw} + \sqrt{\gamma_{c1}} s_2 \tag{S-6-2}$$

where $\Delta\omega = \omega - \omega_0$ is the laser frequency ($\omega$) detuning from the cavity resonance ($\omega_0$). $X_{1(2)}$ are the complex valued backscattering coefficient from CW to CCW mode (CCW to CW mode) [7], controlled by the notch geometry and inter-scatter distance. If each notch scatter is symmetric, then $|X_1| = |X_2|$. Random surface roughness or tilted triangular notches lead to asymmetric backscattering: $|X_1| \neq |X_2|$. The total loss rate for the resonator ($\gamma_t = \gamma_0 + \gamma_{c1} + \gamma_{c2}$) is the sum of intrinsic resonator loss rate ($\gamma_0$ from bending, scattering or linear absorption by mid-gap defect states in silicon) and the coupling loss to the bus (drop) waveguide ($\gamma_{c1(2)}$). As the asymmetric scatter induce different loss in CW and CCW mode, $\gamma_{0\_CW(CCW)}$ and $\gamma_{t\_CW(CCW)}$ are noted as intrinsic loss rate and total loss rate for CW(CCW) mode [S6-S7]. It is noted that $\gamma_{0\_CW} = \gamma_{0\_CCW}$ and $\gamma_{t\_CW} = \gamma_{t\_CCW}$ for symmetric notch scatters. $s_{1(2)}$ is the excitation field amplitude injected from port 1(2). The reflection field amplitudes collected on port 1 and 2 can be expressed as $s_1' = -s_2 + \sqrt{\gamma_{c1}} a_{ccw}$ and $s_2' = -s_1 + \sqrt{\gamma_{c1}} a_{cw}$. The transmission field amplitudes on port 3 and 4 are $s_3 = \sqrt{\gamma_{c2}} a_{cw}$ and $s_4 = \sqrt{\gamma_{c2}} a_{ccw}$, respectively.

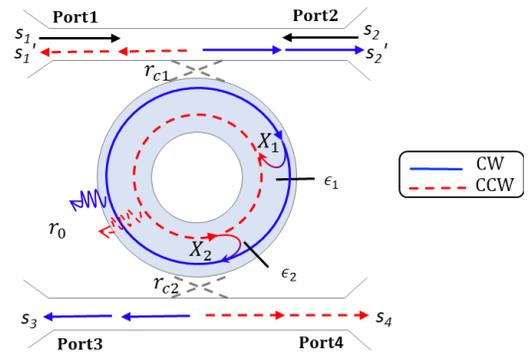

**Fig. S5. Illustration for the coupled harmonic oscillator model.** CW: Clockwise propagating mode; CCW: counterclockwise propagating mode.

The transmittance and reflectance spectra can be derived by solving equation (S-1) in steady state condition, by substituting $\frac{da_{cw(ccw)}}{dt} = 0$. With port 1 excitation ($s_2 = 0$), the normalized transmission spectra ($T$) on through port 2 and 3, and the reflection spectra on port 4 ($R$) can be derived as:

$$T_{1\to 3} = \left|\frac{s_3}{s_1}\right|^2 = \left|\frac{\sqrt{\gamma_{c1}\gamma_{c2}}}{\frac{X_{12}X_{21}}{i\Delta\omega-\gamma_{t\_ccw}/2}+i\Delta\omega-\gamma_{t\_cw}/2}\right|^2 \quad \text{(S-7-1)}$$

$$R_{1\to 4} = \left|\frac{s_4}{s_1}\right|^2 = \left|\frac{-i\sqrt{\gamma_{c1}\gamma_{c2}}X_{12}}{X_{12}X_{21}+(i\Delta\omega-\gamma_{t\_cw}/2)(i\Delta\omega-\gamma_{t\_ccw}/2)}\right|^2 \quad \text{(S-7-2)}$$

With port 2 excitation ($s_1 = 0$), the transmission and reflection spectra can be derived as:

$$T_{2\to 4} = \left|\frac{s_4}{s_2}\right|^2 = \left|\frac{\sqrt{\gamma_{c1}\gamma_{c2}}}{\frac{X_{12}X_{21}}{i\Delta\omega-\gamma_{t\_cw}/2}+i\Delta\omega-\gamma_{t\_ccw}/2}\right|^2 \quad \text{(S-8-1)}$$

$$R_{2\to 3} = \left|\frac{s_3}{s_2}\right|^2 = \left|\frac{-i\sqrt{\gamma_{c1}\gamma_{c2}}X_{21}}{X_{12}X_{21}+(i\Delta\omega-\gamma_{t\_cw}/2)(i\Delta\omega-\gamma_{t\_ccw}/2)}\right|^2 \quad \text{(S-8-2)}$$

Assuming $\gamma_{t\_CW}$ is close to $\gamma_{t\_CCW}$ in the resonator, the transmission spectra are symmetric ($T_{1\to 3} = T_{2\to 4}$) and $\gamma_t$ can be estimated by fitting the transmission spectra.

To minimize the uncertainties caused by grating coupler loss estimation, we compare the intensity ratios between output ports to obtain the best estimations of $\chi_1$ and $\chi_2$:

$$\frac{R_{1\to 4}}{T_{1\to 3}} = \left|\frac{i\chi_{12}}{i\Delta\omega-\gamma_{t\_ccw}/2}\right|^2 \quad \text{(S-9-1)}$$

$$\frac{R_{1\to 4}}{R_{2\to 3}} = \left|\frac{\chi_{12}}{\chi_{21}}\right|^2 \quad \text{(S-9-2)}$$

From equation S-9-1, the value of $\chi_1$ can be estimated at detuning $\Delta\omega = 0$:

$$|\chi_{12}| = \frac{\gamma_t}{2}\sqrt{\frac{R_{1\to 4}}{T_{1\to 3}}} \quad \text{(S-10)}$$

We verify the impact of $\chi_1$ phase on the measured transmission and reflection (Fig. S6). We select three examples of $<\chi_{12} = 0°, \pm 28°, \pm 56°$, and plot the correspondent transmission (solid curves) and reflection (dashed curves) spectra based on equation S-7 (black) and S-8 (grey in Fig. S6a-c). Fig. S6d plots extracted data from the simulated spectra (Fig. S6a-c) and compared to the estimated value in equation S-10. With small phase, the empirical quality factor decreases with the measured reflectivity to transmission ratio, however, the trend might be inversed at large angle. It can be seen from Fig. S6c that the one of the split modes is much higher than the other, than thus the FWHM only encounters one of the split modes. The backscattering coefficient dependent transmission and reflection peaks are plotted in Fig. S6f. *It is noted that the transmission spectra in Fig. 4 are highly symmetric, and thus the $<X_1$ is estimated to be between $\pm 10°$ in those fabricated devices.*

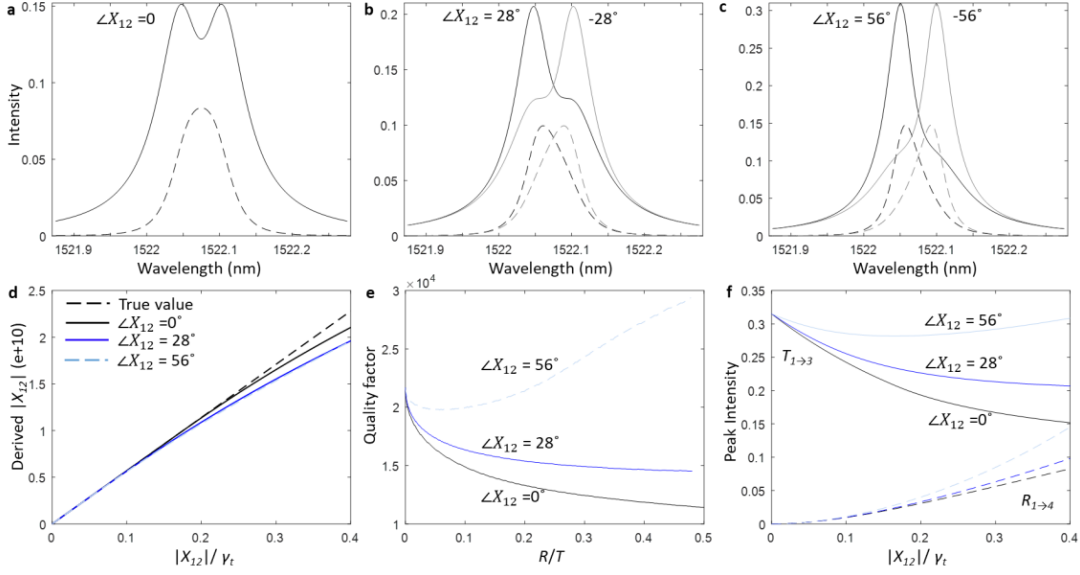

**Fig. S6. $\chi_{12}$ phase dependent spectra**. (a-c) Transmission (solid curve) and reflection (dashed curve) dependent on $\chi_1$ phase; (a) $<\chi_{12}=0°$, (b) $<\chi_{12}=28°$, (c) $<\chi_{12}=56°$. (d) Derived $\chi_{12}$ versus the magnitude of $\chi_1$ with different phase of $\chi_1$, where $\chi_1$ is derived based on Equation (S-10). (e) Quality factor ($=\frac{\lambda_o}{FWHM}$) versus the ratio of the reflection and transmission spectrum as increasing the magnitude of $\chi_1$. (f) Peak intensity versus the magnitude of $\chi_{12}$.

## Supplementary Section 4: Interpretation of the experimental data

For the completeness of the work, we show the transmission and reflection spectra for a set of MRRs with the same radius and notch depth, but varying notch width (*W*) (Fig. S7). We extracted the resonance wavelengths of four adjacent modes of a set of MRRs fabricated in the same run (Fig. S7a). Note that the resonance wavelength variations between devices is due to variations and imperfections in nanofabrication in a university cleanroom. The fabrication related resonance variation is reflected by all four modes as shown in Fig. S7a (e.g. *W* = 100 nm). We see the deterministic control of resonance wavelength (diagonal term in the Hamiltonian) for all four modes in the measured spectral range. We selected the first set of the resonance and analyzed the notch width dependent FWHM and empirical quality factors (Fig. S7b-c). The full transmission and corresponding reflection spectra are given in Fig. S7d-e.

**Experimental loss rates in MRR:** The empirical total loss is estimated from the linewidth of the device (*W* = 200 nm), where the mode-splitting is completely vanished. Estimated total loss ($\gamma_t$) was 62.97 GHz ($Q_t$ ~19.59k). The intrinsic loss rate ($\gamma_o$) and coupling loss rates ($\gamma_{c1}=\gamma_{c2}$) are extracted by fitting the CMT model (equation S-6) into the experimentally measured transmission and reflection spectra. The nanofabrication limited radiation loss $\gamma_o$ is estimated to be 13.321 GHz ($Q_{in}$ = 78.36k). The coupling losses are $\gamma_{c1} = \gamma_{c2} = 39.962$ GHz ($Q_c$ = 52.24k). Those values are applied for fitting the MRRs with varying *W* in Fig. 4a.

**Mie scatterer width dependent backscattering coefficient:** For the MRR with surface roughness only, we extract the $r_{1b} = 19.6754e^{j(-1.1381)}$ GHz and $r_{1f} = 17.5753e^{-j2.5381}$ GHz from the CMT fitting of the selected mode. By analyzing the spectra of the MRRs with varying Mie scatterer design, a constant *Δφ* (~ -0.265337radians) per incremental width (*ΔW* = 40 nm) of Mie scatterer

S-9

is extracted. The result is consistent with our previous study [S8]. The CMT models in Fig. 4a reflect the geometry controlled complex backscattering coefficients.

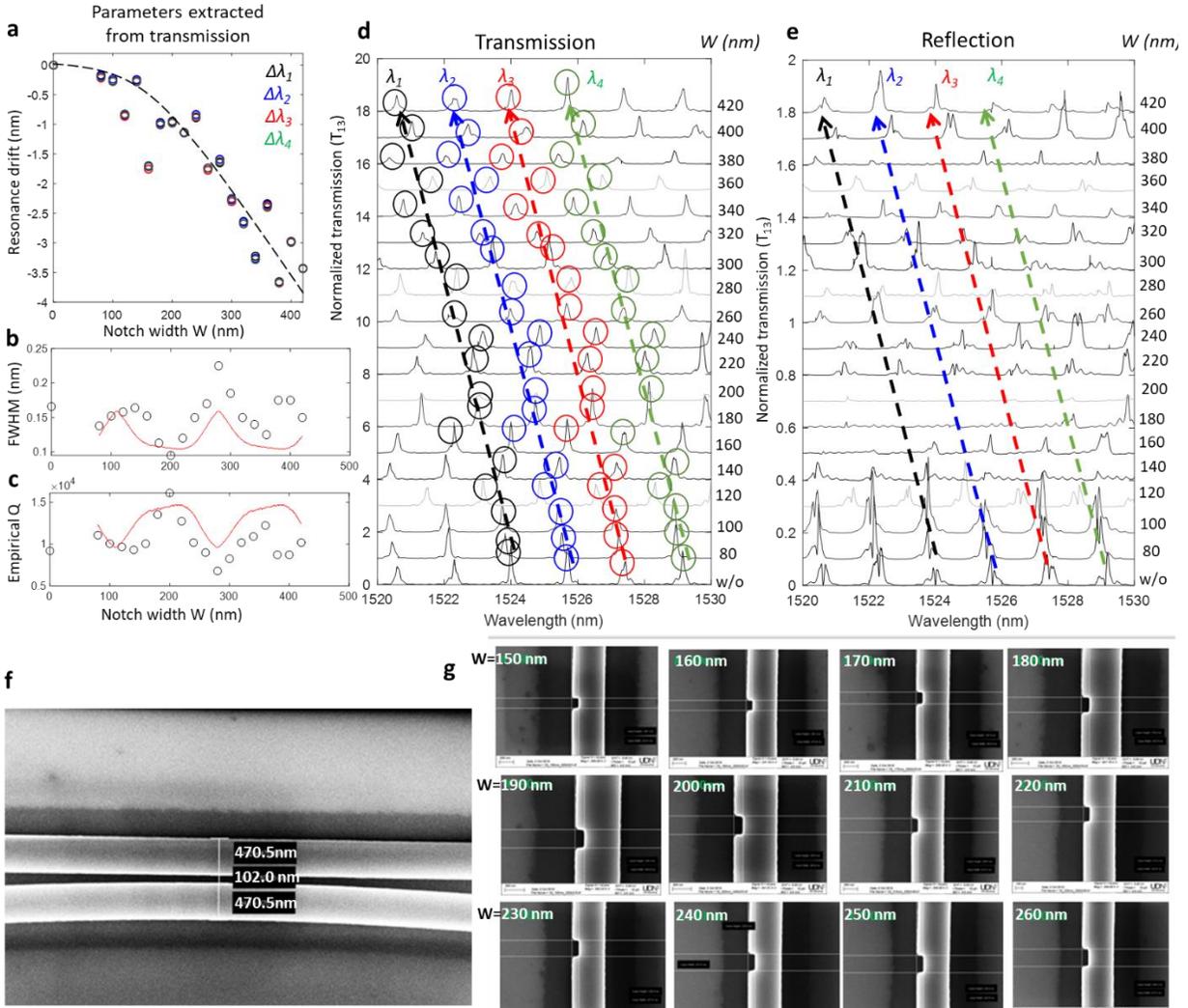

**Figure S7. Measurement results of the set of MRRs with varying notch width.** All the MRRs are fabricated on the same chip. Notch depths are fixed at 100 nm, but notch widths vary in the range 80-400 nm with increments of $\Delta W$ = 20 nm. (a) Notch width dependent resonance wavelength detuning for four adjacent modes, extracted from the transmission spectra in d. (b) Extracted FWHM and (c) empirical quality factor of the MRR, from the transmission spectra for the first mode. (d) Transmission and (e) reflection spectra of the set of MRRs. Offset of 1 is introduced in the transmission spectra and 0.1 for reflection spectra for clarity. (f) Scanning electron microscope image of the ring-waveguide region (fixed for all the MRRs) and (g) Mie scatterers with different widths (W) for the results shown in d-e.

**Supplementary Section 5: Comparison with other works**

Table S1 provides a comparison of our device with other reported schemes for backscattering suppression effect. Suppression of backscattering has been demonstrated in whispering gallery mode resonators (WGMRs) [S9-S11] and topological photonic edge states [S12-S15]. The topological edge state relies on designed topology to achieve directional transport, with assistance of external applied magnetic field. In WGMRs, the suppression is achieved either by introducing scanning nanotip(s) into the mode volume of the resonator [S9, S10] or by utilizing time-reversal symmetry breaking using Brillouin scattering interactions [S11]. Ref [S9] uses an on-chip silica microtoroid resonator, Refs [S10] and [S11] utilizes silica rod and silica microsphere, respectively (These are not on-chip structures). Tapered fibers are used coupling light in and out of the resonators with coupling strength controlled by placing either the resonator or the tapered fiber on a piezo stage. As such, these systems do not lend themselves for building practical devices. The concepts should be translated into completely on-chip dynamically controlled structures. The device demonstrated in this paper addresses this issue. Our device is a silicon MRR fabricated in add-drop configurations with two side-coupled waveguides (i.e., fiber tapered coupling is replaced with on-chip silicon waveguides with gratings). Instead of scanning nanotips introduced into the mode volume of the resonator [S9, S10], we fabricate lithographically defined symmetric and/or asymmetric Mie scatterers to suppress backscattering and bring the system to an EP.

Table S1: Nano/micro-photonic structures for back reflection suppression.

| Device structure | Suppression mechanism | Suppression ratio | Notes | Footprint | $Q$ |
|---|---|---|---|---|---|
| Silica microtoroid [S9] | Tuning silica nanotips positions by piezo stage | ~20 dB* | Mechanically fragile / not scalable/require post tuning | $10^{-8}$ m$^2$ | $>10^7$ |
| Silica rod [S10] | Tuning tungsten nanotip position by piezo stage | 34 dB | | $10^{-6}$ m$^2$ (bulk) | $>10^8$ |
| Silica microsphere [S11] | Tuning the Brillion scattering | NA | | $10^{-8}$ m$^2$ (bulk) | $>10^7$ |
| Photonic crystal edge state [S12-S15] | Lithographically pre-defined topological photonic interface | < 10 dB* [S15] | Stable / scalable / predefined | $10^{-10}$ m$^2$ | $10^2$ [S14] |
| MRR with single Mie scatterer [this work] | Lithographically pre-designed subwavelength defect geometry | 21.41 dB | Stable / scalable / predefined | $10^{-10}$ m$^2$ | $10^5$ |

* Values estimated from the given spectra with forward and backward excitations.